\documentstyle[eqsecnum,prd,preprint,aps]{revtex}
\tighten
\begin{document}
\baselineskip=14pt

\draft

\title{Cosmic Microwave Background anisotropies from second order
gravitational perturbations} 
\author{Silvia Mollerach$^{1}$ and  Sabino Matarrese$^{2}$}
\address{$^{1}$Departamento de Astronomia y Astrof\'\i sica, 
Universidad de Valencia, E-46100 Burjassot, Valencia, Spain}
\address{$^{2}$Dipartimento di Fisica ``G. Galilei'', Universit\`a di
Padova, via Marzolo 8, 35131 Padova, Italy} 

\maketitle

\begin{abstract}
This paper presents a complete analysis of the effects of second 
order gravitational perturbations on Cosmic Microwave Background 
anisotropies, taking explicitly into account scalar, vector and 
tensor modes. We also consider the second order perturbations of
the metric itself obtaining them, for a universe dominated by a 
collision-less fluid, in the Poisson gauge, by transforming 
the known results in the synchronous gauge. We discuss the resulting 
second order anisotropies in the Poisson gauge, and analyse the
possible relevance of the different terms. We expect that, in the 
simplest scenarios for structure formation, the main effect comes
from the gravitational lensing by scalar perturbations, that 
is known to give a few percent contribution to the anisotropies at 
small angular scales.
 
\end{abstract}
\pacs{98.79.Vc,04.25.Nx,98.80.-k}


\section{Introduction}

\setcounter{equation}{0}

The increasing number of measurements of Cosmic Microwave Background 
(CMB) anisotropies in the last years and the very ambitious 
observational programs for the future generation of detectors 
makes us hope that the angular spectrum of the anisotropies will be 
known with great accuracy within the next decade. This fact has 
stimulated theoretical efforts to obtain more precise predictions 
for the anisotropies produced in the different structure formation 
models, and it is expected that future observations will be very 
helpful in distinguishing among them and in putting constraints on the 
cosmological parameters. 

Most of these theoretical computations involve
numerical or semi-analytic solutions of the linearized 
Boltzmann equation. Non-linear gravitational effects on the 
anisotropies have been computed for some particular processes, such as
the gravitational lensing from density perturbations 
\cite{bl87,co89,to89,li90,ca93,mu96} and the Rees-Sciama effect 
\cite{rs68,ma90,ma92,ma94,ar94,tu95}
(which is second order in a flat matter dominated 
universe, as the gravitational potential is constant to first order).
It has been shown that the effect of the gravitational lensing by 
density perturbations is to smooth the so called Doppler or acoustic 
peaks in the angular spectrum at high $\ell$, and it is thus relevant 
for the analysis of the small angle observations \cite{se96b}. 
On the other hand, the non-linear 
Rees-Sciama effect is in most cases expected to be 
much smaller than the first order anisotropies \cite{se96a,sa96} 
unless early reionization substantially erases the first order 
anisotropies. 

In a recent paper, Pyne and Carroll \cite{py96} have presented a 
nice framework for a complete computation of second and higher order
gravitational perturbations of the CMB. Their algorithm essentially
involves computing the redshift experienced by the photons during 
their travel from the last scattering to the observer in terms of 
their perturbed geodesics and then obtaining the perturbed geodesics
up to the required order. The study of second order anisotropies 
is relevant because they can produce a non negligible contribution
compared to the first order ones, due to the long distances involved in 
the problem. The reason is that several second order terms include 
integrals of the metric perturbations along the photons path,
that can enhance small effects as the photons travel from
the last scattering surface. Moreover, second order effects are also 
important because they give the primary contribution to some 
statistical measures of the anisotropies that are vanishing for 
the linear contribution, as for example the three--point
function of temperature anisotropies \cite{lu93,mo95,mu95}.
 In any case, it is important to know the magnitude of the second 
order effects as they contribute to the theoretical error of linear 
anisotropy calculations.

In this paper, we apply the formalism proposed by Pyne and Carroll to
the computation of the full second order anisotropies in the Poisson
gauge. We first present a computation of the second order anisotropies
that generalizes the results of ref. \cite{py96}, in that we consider 
the motion of the observer and the emitter, we explicitly include the
second order perturbations of the metric itself and we take into account
scalar, vector and tensor modes. We then consider the 
Poisson gauge, that, in the case of scalar first order perturbations, 
reduces to the longitudinal gauge. We obtain the second order 
perturbed metric for a dust dominated universe in the Poisson gauge 
explicitly, and then discuss the second order anisotropies for this 
particular case.

Throughout this paper Greek indices
$\mu,\nu,\ldots$ take values from 0 to 3, and Latin ones
$i,j,\ldots$ from 1 to 3. We take, for the metric, signature +2; 
units are such that $c=1$.

\section{Temperature anisotropies}

The quantity of interest is the angular variation of the temperature
measured by an observer.

We consider a perturbed flat Robertson-Walker space-time and use 
conformal time $\eta$ ($d\eta \equiv dt/a(t)$, where $a(t)$ is the 
scale factor of the universe). We can write the line element as
\begin{equation}
ds^2\simeq a^2(\eta)\left(g_{\mu\nu}^{(0)}+g_{\mu\nu}^{(1)}+
\frac{1}{2} g_{\mu\nu}^{(2)}+\ldots\right)dx^\mu dx^\nu,
\end{equation}
where the term between brackets is the conformally transformed
metric ($g_{\mu\nu}$), $g_{\mu\nu}^{(0)}$ is the background 
Minkowski metric, and $g_{\mu\nu}^{(1)}$ and $g_{\mu\nu}^{(2)}$
are the first and second order perturbations respectively.

Photons travel along null geodesics $x^\mu(\lambda)$, that we
parametrize with $\lambda$ in the conformal metric, connecting 
the observer, at coordinates $x^\mu_{\cal{O}}=(\eta_{\cal{O}},
{\bf x}_{\cal{O}})$, to the emitting hypersurface, that we take at 
constant conformal time $\eta_{\cal{E}}$. This hypersurface can be 
taken as the last scattering surface, at redshift $z_{LS}$. At
larger redshifts the hydrogen is ionized and Compton scattering 
off electrons (linked to photons by electromagnetic interactions)
couples photons to baryons. At $z_{LS}$, hydrogen recombines and 
photons can start their travel. We assume that thermal radiation 
with temperature $T_{\cal{E}}({\bf p},\hat{\bf d})$ is emitted by every
point with coordinates $p_i$ in this hypersurface. This temperature
depends also on the direction of emission described by the vector
$\hat{\bf d}$, normalized to unity in the background. The different 
photon paths are specified by the direction from which they arrive 
at $\cal{O}$, specified by a vector $\hat{\bf e}$ normalized to unity
in the background.
This direction fixes the point ${\bf p}$ and the direction 
$\hat{\bf d}$ at emission.

If the CMB has a black body distribution and the photons suffer a 
redshift $z$ during their travel from the emitter $\cal{E}$ to the 
observer $\cal{O}$, the emitted frequency $\omega_{\cal{E}}$ and the 
observed one $\omega_{\cal{O}}$ are related by $\omega_{\cal{O}}=
\omega_{\cal{E}}/(1+z)$. Since the occupation number per frequency mode
is conserved, the corresponding photon temperatures are related 
by $T_{\cal{O}}= T_{\cal{E}}/(1+z)$. The anisotropies detected by an 
observer are  due to inhomogeneities in the temperature at emission and
to the different redshift suffered by photons coming from different
directions. We will compute this quantity up to second order in
gravitational perturbations.

The temperature 
measured by an observer at  $\cal{O}$ can be written as
\begin{equation}
T_{\cal{O}} ({\bf x}_{\cal{O}},\hat{\bf e})=
\frac{\omega_{\cal{O}}}{\omega_{\cal{E}}} 
T_{\cal{E}}({\bf p},\hat{\bf d}),
\label{tempo}
\end{equation}
with $\omega=-g_{\mu\nu} U^\mu k^\nu$, where $U^\mu$ is the 
four-velocity of the observer or emitter and $k^\nu=dx^\nu/d\lambda$
is the wave vector of the photon in the conformal metric, tangent to 
the null geodesic $x^\nu(\lambda)$ followed by the photon from the
emission to the observation point. In fact, we will propagate photons 
back from the observation point to the emission surface. 
We thus need to obtain 
$\omega_{\cal{E}}$, ${\bf p}$ and $\hat{\bf d}$ for a given initial set 
of values ${\bf x}_{\cal{O}}$, $\hat{\bf e}$ and $\omega_{\cal{O}}$. The 
resulting quantities are functions of the photon path and wave vector,
that we expand in series of the metric perturbations $g_{\mu\nu}^{(r)}$
and their derivatives as
\begin{eqnarray}
x^\mu(\lambda)&=&x^{(0)\mu}(\lambda)+x^{(1)\mu}(\lambda)+
x^{(2)\mu}(\lambda)+\ldots , \nonumber\\
k^\mu(\lambda)&=&k^{(0)\mu}(\lambda)+k^{(1)\mu}(\lambda)+
k^{(2)\mu}(\lambda)+\ldots .
\end{eqnarray}

Contrary to the assumptions of ref. \cite{py96}, we are not taking the 
observer and emitter comoving with the total fluid of the universe. 
In this way we keep track, to the first order, of the dipole due 
to the observer's motion and of the Doppler effect due to the 
emitter's motion, that are otherwise lost. 
Besides these effects, to the second order, we also take into 
account cross terms involving the velocities and other sources of 
anisotropy.

We can expand the four-velocity as
\begin{equation}
U^\mu=\frac{1}{a}\left(\delta^\mu_0 + v^{(1)\mu}+\frac{1}{2}
 v^{(2)\mu}+\ldots \right).
\end{equation}
This is subject to the normalization condition $U^\mu
U_\mu=-1$.

It is also useful to write the perturbed spatially flat conformal 
metric as 
\begin{equation}\label{eq:m1}
g_{00}=-\left(1+2\psi^{(1)}+\psi^{(2)}+\ldots ,\right)\;,
\end{equation}
\begin{equation}\label{eq:m2}
g_{0i}=z_i^{(1)}+\frac{1}{2} z_i^{(2)}+\ldots ,
\end{equation}
\begin{equation}\label{eq:m3}
g_{ij}=\left(1-2 \phi^{(1)} -\phi^{(2)}\right)\delta_{ij}+
\chi^{(1)}_{ij}+\frac{1}{2}\chi^{(2)}_{ij}+\ldots ,
\end{equation}
where\footnote{Indices are raised and lowered
using $\delta^{ij}$ and $\delta_{ij}$, respectively.}
$\chi^{(r)i}_{i}=0$ and the functions $\psi^{(r)}$, $z^{(r)}_i$,
$\phi^{(r)}$, and $\chi^{(r)}_{ij}$ represent the
 $r$-th order perturbation of the metric.

The normalization condition for the velocity fixes 
the time component $v^{(r)0}$ in terms of 
the lapse perturbation, $\psi^{(r)}$.  For the first and
 second order perturbations we obtain:
\begin{eqnarray}
\label{eq:v0psi1}
v^{(1)0} &= &-\psi^{(1)}\; ; \\
\label{eq:v0psi2}
v^{(2)0} &= &-\psi^{(2)}+3\left(\psi^{(1)}\right)^2
+2 z^{(1)}_i v^{(1)i}+v^{(1)}_iv^{(1)i}\; .
\end{eqnarray}

In order to obtain the variation in the sky of the observed temperature
up to the second order, according to eq. (\ref{tempo}), we need to 
expand $\omega_{\cal{O}}$ and $\omega_{\cal{E}}$ up to second order in 
gravitational perturbations
\begin{equation}
\omega=\omega^{(0)}\left(1+\tilde{\omega}^{(1)}+
\tilde{\omega}^{(2)}+\ldots \right),
\end{equation}
and also to expand the temperature at emission 
\begin{equation}
T_{\cal{E}}({\bf p},\hat{\bf d})=T_{\cal{E}}^{(0)}\left(1+
\tau ({\bf p},\hat{\bf d}) \right).
\end{equation}
We will not perform a full expansion of $\tau ({\bf p},\hat{\bf d})$, 
as a calculation of this quantity would be beyond the aim of this 
paper. We will instead assume that it is known for a given model and 
compute the additional effect of gravity along the photons  path.
We also have to take into account that the point  ${\bf p}$ and 
direction $\hat{\bf d}$ at emission need to be expanded in the 
expression of $\tau ({\bf p},\hat{\bf d})$ as 
 ${\bf p}={\bf p}^{(0)}+{\bf p}^{(1)}+\ldots$, and 
 $\hat{\bf d}={\bf d}^{(0)}+{\bf d}^{(1)}+\ldots$. 
Performing these expansions
in eq. (\ref{tempo}) we obtain \cite{py96}
\begin{eqnarray}
T_{\cal{O}} ({\bf x}_{\cal{O}},\hat{\bf e})&=&
\frac{\omega_{\cal{O}}^{(0)}}{\omega_{\cal{E}}^{(0)}} 
T_{\cal{E}}^{(0)}
\left[1+\left(\tilde{\omega}_{\cal{O}}^{(1)}
-\tilde{\omega}_{\cal{E}}^{(1)}+\tau\right)\right.\nonumber\\&+&
\left(\tilde{\omega}_{\cal{O}}^{(2)}
-\tilde{\omega}_{\cal{E}}^{(2)}
+(\tilde{\omega}_{\cal{E}}^{(1)})^2-\tilde{\omega}_{\cal{O}}^{(1)}
\tilde{\omega}_{\cal{E}}^{(1)}+\tilde{\omega}_{\cal{O}}^{(1)}\tau
 -\tilde{\omega}_{\cal{E}}^{(1)}\tau\right.\nonumber\\&
+&\left.\left. p^{(1)i}\frac{\partial \tau}{\partial x^i}
+d^{(1)i}\frac{\partial \tau}{\partial d^i}\right)+\ldots
\right]
\end{eqnarray}
where $\tau$ and its spatial derivatives have to be evaluated at
(${\bf p}^{(0)},{\bf d}^{(0)}$). The first factor gives the mean 
temperature at the observation point $T_{\cal{O}}^{(0)}\equiv
T_{\cal{E}}^{(0)} \omega_{\cal{O}}^{(0)}/\omega_{\cal{E}}^{(0)}$, and
the round brackets inside the term in square brackets define the first 
and second order perturbations, $\delta T^{(1)}$ and $\delta T^{(2)}$, 
that we are looking for.

To compute them, we will use the same background geodesics as in ref.
\cite{py96}
\begin{eqnarray}
x^{(0)\mu}&=&(\lambda,(\lambda_{\cal{O}} -\lambda) e^i),\nonumber\\
k^{(0)\mu}&=&(1,- e^i),
\end{eqnarray}
and boundary conditions at the origin
\begin{eqnarray}
x^{(1)\mu}(\lambda_{\cal{O}})&=&x^{(2)\mu}(\lambda_{\cal{O}})=0
,\nonumber\\
k^{(1)i}(\lambda_{\cal{O}})&=&k^{(2)i}(\lambda_{\cal{O}})=0.
\end{eqnarray}
The condition that the wave vector is null fixes the value of
$k^{(1)0}(\lambda_{\cal{O}})$ and $k^{(2)0}(\lambda_{\cal{O}})$.
We will only need $k^{(1)0}(\lambda_{\cal{O}})$ explicitly
\begin{equation}
k^{(1)0}(\lambda_{\cal{O}})=-\psi^{(1)}_{\cal{O}}
-z^{(1)i}_{\cal{O}} e_i -\phi^{(1)}_{\cal{O}}+\frac{1}{2}
\chi^{(1)ij}_{\cal{O}}e_i e_j.
\end{equation}

Using the metric, four-velocity and wave vector expansions we can
obtain the quantities in the expansion of $\omega$
\begin{eqnarray}
\omega^{(0)}&=&a^{-1},\nonumber\\
\tilde{\omega}^{(1)}&=&k^{(1)0}+\psi^{(1)}+v^{(1)}_i e^i
+z^{(1)}_i e^i,\nonumber\\
\tilde{\omega}^{(2)}&=&k^{(2)0}+\frac{1}{2}\psi^{(2)}+
\frac{1}{2}z^{(2)}_i e^i+\frac{1}{2}v^{(2)}_i e^i
-\frac{1}{2}(\psi^{(1)})^2+\frac{1}{2}v^{(1)}_i v^{(1)i}
+k^{(1)0}\psi^{(1)}-v^{(1)}_i k^{(1)i}-z^{(1)}_i k^{(1)i}\nonumber\\
&-&\psi^{(1)}z^{(1)}_i e^i -2\phi^{(1)}v^{(1)}_i e^i
+\chi^{(1)}_{ij}e^i v^{(1)j}+\frac{d k^{(1)0}}{d\lambda}
\Delta\lambda+(\psi^{(1)}_{,j}+z^{(1)}_{i,j}e^i+v^{(1)}_{i,j}e^i) 
p^{(1)j},
\end{eqnarray}
where $\Delta\lambda$ is the difference in affine parameter between 
the points where the background and first order geodesics intersect
the $\eta=\eta_{\cal{E}}$ hypersurface, 
and is given by  $\Delta\lambda=-x^{(1)0}+\ldots$.
It can also be seen \cite{py96} that $p^{(1)i}=x^{(1)i}+x^{(1)0} e^i$
and
\begin{equation}
d^{(1)i}=e^i-\frac{e^i-k^{(1)i}}{|e^i-k^{(1)i}|}.
\end{equation}

Finally, we obtain, for the first order temperature anisotropy, 
\begin{eqnarray}
\delta T^{(1)}&=&\tilde{\omega}^{(1)}_{\cal{O}}-
\tilde{\omega}^{(1)}_{\cal{E}}+\tau\nonumber\\
&=&-\phi^{(1)}_{\cal{O}}+\frac{1}{2}\chi^{(1)ij}_{\cal{O}}e_i e_j+
v^{(1)i}_{\cal{O}} e_i-k^{(1)0}_{\cal{E}}-v^{(1)i}_{\cal{E}} e_i
-z^{(1)i}_{\cal{E}} e_i-\psi^{(1)}_{\cal{E}}+\tau,
\label{dt1}
\end{eqnarray}
and, for the second order one, 
\begin{eqnarray}
\delta T^{(2)}&=&\left(k^{(2)0}+\frac{1}{2}\psi^{(2)}+
\frac{1}{2}v^{(2)}_i e^i+\frac{1}{2}z^{(2)}_i e^i
-\frac{1}{2}(\psi^{(1)})^2+\frac{1}{2}v^{(1)}_i v^{(1)i}
+k^{(1)0}\psi^{(1)}\right.\nonumber\\
&-&\left.\left.\psi^{(1)}z^{(1)}_i e^i -2\phi^{(1)}v^{(1)}_i e^i
+\chi^{(1)}_{ij}e^i v^{(1)j}\right)\right|^{\cal{O}}_{\cal{E}}
+(v^{(1)}_{{\cal{E}}i}+z^{(1)}_{{\cal{E}}i}) k^{(1)i}_{\cal{E}}
+\left.\frac{d k^{(1)0}}{d\lambda}\right|_{\cal{E}}
x^{(1)0}_{\cal{E}}\nonumber\\
&-&\left(\psi^{(1)}_{,j}+z^{(1)}_{i,j} e^i+v^{(1)}_{i,j} e^i+
\tau_{,j}\right)_{\cal{E}}(x^{(1)j}+x^{(1)0} e^j)_{\cal{E}}
+\left.\frac{\partial \tau}{\partial d^i}\right|_{\cal{E}}
d^{(1)i} \nonumber\\
&-&\left(k^{(1)0}+v^{(1)}_i e^i+z^{(1)}_i e^i+\psi^{(1)}-
\tau\right)_{\cal{E}}\left. \left(k^{(1)0}+v^{(1)}_i e^i
+z^{(1)}_i e^i+\psi^{(1)}\right)\right|^{\cal{O}}_{\cal{E}}.
\label{dt2}
\end{eqnarray}
The next step is to obtain the null geodesics up to second order;
in particular, we need to compute the quantities $k^{(2)0}$, 
$k^{(1)\mu}$ and $x^{(1)\mu}$ to substitute in eqs. (\ref{dt1}) and
(\ref{dt2}). This problem has been solved for a general perturbed
spacetime in ref. \cite{py96} using the geodesic expansion 
introduced by Pyne and Birkinshaw \cite{py93}. Following their method,
we obtain, for perturbations around a flat Robertson-Walker background 
in any gauge, that the first order wave vector is given by
\begin{equation}
k^{(1)0}(\lambda_{\cal{E}})=\psi^{(1)}_{\cal{O}}
-\phi^{(1)}_{\cal{O}}+\frac{1}{2}
\chi^{(1)ij}_{\cal{O}}e_i e_j-2 \psi^{(1)}_{\cal{E}}
-z^{(1)i}_{\cal{E}} e_i +I_1(\lambda_{\cal{E}}),
\label{k10}
\end{equation}
with 
\begin{equation}
I_1(\lambda_{\cal{E}})=\int_{\lambda_{\cal{O}}}^{\lambda_{\cal{E}}}
d\lambda A^{(1)'},
\label{i1}
\end{equation}
where $A^{(1)}\equiv \psi^{(1)}+\phi^{(1)}+z^{(1)}_i e^i-
\frac{1}{2}\chi^{(1)}_{ij}e^i e^j$, and
\begin{equation}
k^{(1)i}(\lambda_{\cal{E}})=2 \phi^{(1)}_{\cal{O}} e^i+
z^{(1)i}_{\cal{O}}-\chi^{(1)ij}_{\cal{O}} e_j
-2 \phi^{(1)}_{\cal{E}} e^i-
z^{(1)i}_{\cal{E}}+\chi^{(1)ij}_{\cal{E}} e_j
- I_1^i(\lambda_{\cal{E}}),
\end{equation}
with
\begin{equation}
I^i_1(\lambda_{\cal{E}})=\int_{\lambda_{\cal{O}}}^{\lambda_{\cal{E}}}
d\lambda A^{(1),i}.
\end{equation}

For the first order geodesics, we obtain 
\begin{eqnarray}
x^{(1)0}(\lambda_{\cal{E}})&=&(\lambda_{\cal{E}}-\lambda_{\cal{O}})
\left[\psi^{(1)}_{\cal{O}}-\phi^{(1)}_{\cal{O}}+\frac{1}{2}
\chi^{(1)ij}_{\cal{O}}e_i e_j\right]+
\int_{\lambda_{\cal{O}}}^{\lambda_{\cal{E}}}d\lambda
\left[-2 \psi^{(1)}-z^{(1)}_i e^i+(\lambda_{\cal{E}}-\lambda)
A^{(1)'}\right],\nonumber\\
x^{(1)i}(\lambda_{\cal{E}})&=&(\lambda_{\cal{E}}-\lambda_{\cal{O}})
\left[2 \phi^{(1)}_{\cal{O}} e^i+
z^{(1)i}_{\cal{O}}-\chi^{(1)ij}_{\cal{O}} e_j\right]\nonumber\\
&-&\int_{\lambda_{\cal{O}}}^{\lambda_{\cal{E}}}d\lambda
\left[2 \phi^{(1)} e^i+z^{(1)i}-\chi^{(1)ij} e_j
+(\lambda_{\cal{E}}-\lambda)A^{(1),i}\right] .
\end{eqnarray}

For the second order, we need only the difference between the wave
vector at emission and observation
\begin{eqnarray}
k^{(2)0}_{\cal{E}}-k^{(2)0}_{\cal{O}}&=&\psi^{(2)}_{\cal{O}}
-\psi^{(2)}_{\cal{E}}-\frac{1}{2}z^{(2)i}_{\cal{E}} e_i+
\frac{1}{2}z^{(2)i}_{\cal{O}} e_i +2 \psi^{(1)}_{\cal{O}}
k^{(1)0}_{\cal{O}}-2 \psi^{(1)}_{\cal{E}}
k^{(1)0}_{\cal{E}}\nonumber\\
&-& \left(2 x^{(1)i}\psi^{(1)}_{,i}+2 x^{(1)0}
\psi^{(1)'}-z^{(1)i}k^{(1)}_i+x^{(1)0} z^{(1)'}_i e^i+
x^{(1)i} z^{(1)}_{j,i} e^j\right)_{\cal{E}} +I_2(\lambda_{\cal{E}}),
\end{eqnarray}
with 
\begin{eqnarray}
I_2(\lambda_{\cal{E}})&=&\int_{\lambda_{\cal{O}}}^{\lambda_{\cal{E}}}
d\lambda \left[\frac{1}{2}A^{(2)'}-(z^{(1)'}_i-\chi^{(1)'}_{ij} e^j)
(k^{(1)i}+e^i k^{(1)0})\right.\nonumber\\
&+&\left. 2 k^{(1)0} A^{(1)'} +2 \phi^{(1)'} A^{(1)}
+x^{(1)0} A^{(1)''}+x^{(1)i} A^{(1)'}_{,i}\right],\label{i2}
\end{eqnarray}
where $A^{(2)}\equiv \psi^{(2)}+\phi^{(2)}+z^{(2)}_i e^i-
\frac{1}{2}\chi^{(2)}_{ij}e^i e^j$.

We can now write the temperature anisotropy in terms of the 
metric perturbations. Replacing eq. (\ref{k10}) into (\ref{dt1})
we obtain, for the first order,
\begin{equation}
\delta T^{(1)}=\psi^{(1)}_{\cal{E}}-\psi^{(1)}_{\cal{O}}+
v^{(1)i}_{\cal{O}} e_i-v^{(1)i}_{{\cal{E}}} e_i+\tau
-I_1(\lambda_{\cal{E}}).\label{dt1p}
\end{equation}
This is a general expression, valid in any gauge, that takes 
into account scalar, vector and tensor perturbations. It 
includes the effect of intrinsic anisotropies in the last 
scattering surface, dipole due to the observer's motion,
Doppler effect from the emitter's motion and gravitational  
redshift of the photons. It is equivalent to the well-known
result originally obtained by Sachs and Wolfe \cite{sa67}.
It can be seen that the full expression is gauge invariant 
up to a monopole term; the relative contributions from the
intrinsic, Doppler and gravitational redshift contributions 
are however gauge dependent.

Analogously, for the second order, we obtain
\begin{eqnarray}
\delta T^{(2)}&=&\frac{1}{2}\psi^{(2)}_{\cal{E}}
-\frac{1}{2}\psi^{(2)}_{\cal{O}}
+\frac{3}{2}(\psi^{(1)}_{\cal{O}})^2
-\frac{1}{2}(\psi_{\cal{E}}^{(1)})^2-I_2(\lambda_{\cal{E}})
-v^{(1)i}_{\cal{E}} e_i \psi^{(1)}_{\cal{E}}\nonumber\\
&+&\left(I_1(\lambda_{\cal{E}})+v^{(1)i}_{\cal{E}} e_i\right)
\left(2 \psi^{(1)}_{\cal{O}}
-\phi^{(1)}_{\cal{O}}+\frac{1}{2}\chi^{(1)ij}_{\cal{O}}e_i e_j
-v^{(1)i}_{\cal{O}} e_i
-\psi^{(1)}_{\cal{E}}-\tau+v^{(1)i}_{\cal{E}} e_i
+I_1(\lambda_{\cal{E}})\right)\nonumber\\
&+&x^{(1)0}_{\cal{E}} A ^{(1)'}_{\cal{E}}+(x^{(1)j}_{\cal{E}}
+x^{(1)0}_{\cal{E}} e^j)\left(\psi^{(1)}_{,j}-v^{(1)}_{i,j} e^i+
\tau_{,j}\right)_{\cal{E}}+v^{(1)i}_{\cal{O}}
\left(\frac{1}{2}v^{(1)}_{{\cal{O}}i}-2 \phi^{(1)}_{\cal{O}} e_i
+\chi^{(1)}_{{\cal{O}}ij} e^j\right)\nonumber\\
&-&\frac{1}{2}v^{(1)}_{{\cal{E}}i} v^{(1)i}_{\cal{E}}
+\psi^{(1)}_{\cal{E}}\tau+
\frac{\partial \tau}{\partial d^i}d^{(1)i}
-\psi^{(1)}_{\cal{O}}(\psi^{(1)}_{\cal{E}}+\tau)\nonumber\\
&-&v^{(1)i}_{\cal{O}} e_i\left(\psi^{(1)}_{\cal{O}}
-\phi^{(1)}_{\cal{O}}+\frac{1}{2}\chi^{(1)kj}_{\cal{O}}e_k e_j
-\tau-\psi^{(1)}_{\cal{E}}\right)
\nonumber\\
&+&v^{(1)}_{{\cal{E}}i}\left(-z^{(1)i}_{\cal{E}}
+z^{(1)i}_{\cal{O}}+2 \phi^{(1)}_{\cal{O}}e^i
-\chi^{(1)ij}_{\cal{O}} e_j- I_1^i(\lambda_{\cal{E}})\right).
\label{dt2p}
\end{eqnarray}
This is also a general expression, that is valid in any gauge and
takes into account scalar, vector and tensor perturbations.
It also includes the effects of the motion of the observer and 
the emitter. In the previous expression we have dropped the terms
proportional to $v^{(2)i}$ as this computation is not
aimed at obtaining $v^i$ at the emission or observation points,
but assumes that they are known quantities. 

To proceed further with the computation, we need to know the 
initial values and the evolution of the perturbations. To solve
this it is necessary to fix a gauge. There are different 
possibilities: the synchronous gauge ($\psi^{(r)}=z^{(r)}=0$) 
turns out to be convenient for many calculations and has been 
widely used for linear anisotropy computations.
 Another choice is the Poisson gauge 
($z_i^{(r),i}={\chi_{ij}}^{(r),j}=0$), recently discussed by 
Bertschinger \cite{be96}, that in the case of scalar perturbations 
reduces to the longitudinal gauge. The latter gauge, in which 
$z_i^{(r)}=\chi_{ij}^{(r)}=0$, 
has become very popular, because the evolution 
equations are most similar to the Newtonian ones, and thus closest
to our classical intuition. All second order temperature anisotropy
calculations have been performed in this gauge. Since the vector and 
tensor modes are set to zero by hand, the longitudinal gauge should 
not be used to study perturbations beyond the linear regime: this is 
because in the nonlinear case the scalar, vector, and tensor modes 
are dynamically coupled and vector and tensor modes cannot be set 
to zero arbitrarily. This could be a problem when studying the 
Rees-Sciama effect that explicitly involves non-linearities in the 
metric perturbations; we will come to this point in section IV.
We will use the Poisson gauge, which overcomes the above limitation 
of the longitudinal gauge, while keeping all its advantages in terms 
of physical interpretation of the results. 

\section{Second order perturbations in General Relativity}

We consider the gravitational instability of irrotational collision-less 
matter in a flat Robertson-Walker background up to second order. 
Different approaches to this problem have been proposed. 
The first solution of the second order relativistic equations has been 
obtained, in the synchronous gauge, in a pioneering work by Tomita 
\cite{to67}. Matarrese, Pantano and Saez \cite{ma94a,ma94b} obtained 
the leading order terms of the expansion, using a different method, 
based on the so-called fluid-flow approach. Salopek, Stewart and 
Croudace \cite{ssc94} used a gradient expansion technique to obtain 
second order metric perturbations; an intrinsic limitation of their 
method is, however, that non-local terms, such as the non-linear 
tensor modes, are lost. Russ et {\it al.} \cite{ru96} recently 
rederived the metric perturbations to second order in the synchronous 
gauge, using a tetrad formalism. We are interested here in obtaining 
the solution in the Poisson gauge. Instead of perturbing the Einstein 
equations in this gauge and then solving them, we will transform the 
solution known in the synchronous gauge to the Poisson one, using the 
second order gauge transformation recently developed in ref. \cite{br96} 
(for more details see ref. \cite{ma97}).

Up to this point we have been completely general in the inclusion of 
scalar, vector and tensor modes. Now, in order to give a more 
quantitative insight on the relevance of the different contributions, 
we will make some restrictions. First, we will neglect vector modes 
at the linear order. The fact that they have decreasing amplitude and 
that they are not generated in inflationary theories, makes us expect 
that their role will not be relevant, at least if inflation was the 
mechanism for primordial fluctuation generation. We will however keep 
track of the second order vector modes generated by the coupling with 
scalar modes. Second, we will neglect the effect of linear tensor modes 
as sources for second order metric perturbations: because of the 
graviton free-streaming inside the horizon this is a very reasonable 
approximation. We will then only consider the second order 
perturbations generated by linear scalar perturbations.
These are expected to give the dominant contribution, as vector and 
tensor modes have decaying amplitude. The solutions given in refs. 
\cite{to67,ru96}, and hence those obtained in this
section apply to this restricted case. For the computation of the 
CMB anisotropies we will however keep the contribution of the linear
tensor modes in $\delta T^{(1)}$ and $\delta T^{(2)}$ everywhere,
except as source for $g_{\mu\nu}^{(2)}$. The contribution of tensor
modes can in fact be comparable to the scalar contribution to 
$\delta T^{(1)}$ at large scales in many inflationary models 
\cite{lu92}.

The first order solution to the perturbed Einstein equations in the 
synchronous gauge is given by (see, e.g., ref. \cite{ks84})
\begin{eqnarray}
\psi_S^{(1)}&=&z_S^{(1)}=0,\nonumber\\
\phi_S^{(1)}&=&\frac{5}{3}\varphi
+\frac{\eta^2}{18}\nabla^2\varphi,\nonumber\\
\chi_{S ij}^{(1)}&=&-\frac{\eta^2}{3}\left(\varphi_{,ij}
-\frac{1}{3}\delta_{ij}\nabla^2\varphi\right)+\chi_{ij}^{\top(1)},
\end{eqnarray}
where $\varphi=\varphi({\bf x})$ is the initial peculiar gravitational
potential. $\chi_{ij}^{\top(1)}$ is the tensor (transverse and
traceless) contribution that can be written as
\begin{equation}
\chi_{ij}^{\top(1)}({\bf x},\eta)=\frac{1}{(2\pi)^3} \int d^3{\bf k}
\exp(i{\bf k}\cdot{\bf x}) \chi^{(1)}_\sigma({\bf k},\eta) 
\epsilon^{\sigma}_{ij}(\hat{\bf k}),
\end{equation}
where $\epsilon^{\sigma}_{ij}(\hat{\bf k})$ is the polarization tensor,
with $\sigma$ ranging over the polarization components $+,\times$, and 
$\chi^{(1)}_\sigma({\bf k},\eta)$ is the amplitude. Its time evolution
during the matter dominated era can be represented as
\begin{equation}
\chi^{(1)}_\sigma({\bf k},\eta) \approx A(k) a_\sigma({\bf k}) 
\left(\frac{3 j_1(k\eta)}{k\eta}\right),
\end{equation}
where $a_\sigma({\bf k})$ is a zero mean random variable with 
auto-correlation function 
$\langle a_\sigma({\bf k}) a_{\sigma'}({\bf k'})\rangle =(2\pi)^3 
k^{-3} \delta^3({\bf k}+{\bf k'}) \delta_{\sigma\sigma'}$.
The spectrum of the gravitational wave background depends on the 
processes by which it was generated, and for example in most 
inflationary models, $A(k)$ is nearly scale invariant and proportional
to the Hubble constant during inflation.

The second order perturbations are given by \cite{to67,ru96,ma97}
\begin{eqnarray}
\psi_S^{(2)}&=&z_S^{(2)}=0,\nonumber\\
\phi_S^{(2)}&=&\frac{\eta^4}{252}\left(-\frac{10}{3}\varphi^{,ki}
\varphi_{,ki}+(\nabla^2\varphi)^2\right)
+\frac{5}{18}\eta^2\left(\varphi^{,k}\varphi_{,k}+
\frac{4}{3}\varphi\nabla^2\varphi\right),\nonumber\\
\chi_{S ij}^{(2)}&=&\frac{\eta^4}{126}\left(19\varphi^{,k}_{,i}
\varphi_{,kj}-12 \varphi_{,ij} \nabla^2\varphi
+4 (\nabla^2\varphi)^2 \delta_{ij}
-\frac{19}{3}\varphi^{,kl}\varphi_{,kl} \delta_{ij}\right)\nonumber\\
&+&\frac{5}{9}\eta^2\left(-6\varphi_{,i}\varphi_{,j}
-4\varphi \varphi_{,ij}+2 \varphi^{,k}\varphi_{,k}\delta_{ij}
+\frac{4}{3}\varphi\nabla^2\varphi\delta_{ij}\right)
+\pi_{S ij},
\end{eqnarray}
where the traceless and transverse contribution $\pi_{S ij}$ 
satisfies the inhomogeneous wave equation 
\begin{equation}
\pi_{S ij}''+\frac{4}{\eta}\pi_{S ij}'-\nabla^2 \pi_{S ij}=
-\frac{\eta^4}{21}\nabla^2 S_{ij}, 
\end{equation}
with 
\begin{equation}
S_{ij}=\nabla^2 \Psi_0 \delta_{ij}+ \Psi_{0,ij}+
2\left(\varphi_{,ij}\nabla^2\varphi-\varphi_{,ik}
\varphi^{,k}_{,j}\right),
\end{equation}
where
\begin{equation}
\nabla^2 \Psi_0=-\frac{1}{2}\left((\nabla^2\varphi)^2-
\varphi_{,ik}\varphi^{,ik}\right).
\end{equation}
This equation can be solved using the Green method; 
we obtain for $\pi_{S ij}$ that
\begin{equation}
\pi_{S ij}=\frac{\eta^4}{21}S_{ij}+\frac{4}{3}\eta^2 T_{ij}({\bf x})
+\tilde{\pi}_{ij}({\bf x},\eta),
\end{equation}
where $\nabla^2 T_{ij}=S_{ij}$ and the remaining piece 
$\tilde{\pi}_{ij}({\bf x},\eta)$, accounting for a term that is 
constant in time and another one that oscillates with decreasing 
amplitude, can be written as
\begin{equation}
\tilde{\pi}_{ij}({\bf x},\eta)=\frac{1}{(2\pi)^3} \int d^3{\bf k}
\exp(i{\bf k}\cdot{\bf x})\frac{40}{k^4}S_{ij}({\bf k})\left(
\frac{1}{3}+\frac{\cos(k\eta)}{(k\eta)^2}-
\frac{\sin(k\eta)}{(k\eta)^3}\right),
\end{equation}
with $S_{ij}({\bf k})=\int d^3{\bf x}\exp(-i{\bf k}\cdot{\bf x})
S_{ij}({\bf x})$.

The gauge transformation is determined to each order by a 
four-vector $\xi^{(r)\mu}$ that we split as $\xi^{(r)0}=
\alpha^{(r)}$ and $\xi^{(r)i}= \partial^i \beta^{(r)}+d^{(r)i}$,
with $\partial_i d^{(r)i}=0$. In ref. \cite{br96} the vectors
$\xi^{(1)\mu}$ and $\xi^{(2)\mu}$, describing the gauge 
transformation from the synchronous to the Poisson gauge,
have been explicitly obtained in terms of the synchronous
metric perturbations $g_{S\mu\nu}^{(1)}$ and $g_{S\mu\nu}^{(2)}$.
Using the metric perturbations in the synchronous gauge presented 
above, we can write the first order gauge transformation as
\begin{eqnarray}
\alpha^{(1)}&=&\frac{\eta}{3} \varphi,\nonumber\\
\beta^{(1)}&=&\frac{\eta^2}{6} \varphi,
\end{eqnarray}
and $d^{(1)i}=0$, in the absence of vector modes in the initial 
conditions. 
For the second order, we obtain
\begin{eqnarray}
\alpha^{(2)}&=&-\frac{2}{21} \eta^3 \Psi_0+\eta
\left(\frac{10}{9} \varphi^2 +4 \Theta_0\right),\nonumber\\
\beta^{(2)}&=&\eta^4\left(\frac{1}{72}\varphi^{,i}\varphi_{,i}
-\frac{1}{42}\Psi_0\right)+\frac{\eta^2}{3}\left(\frac{7}{2} 
\varphi^2+6\Theta_0\right),
\end{eqnarray}
where $\nabla^2 \Theta_0= \Psi_0-\frac{1}{3}\varphi^{,i}
\varphi_{,i}$ and
\begin{equation}
\nabla^2 d^{(2)}_j=\eta^2\left(-\frac{4}{3}\varphi_{,j}
\nabla^2\varphi+\frac{4}{3}\varphi^{,i}\varphi_{,ij}
-\frac{8}{3}\Psi_{0,j}\right).
\end{equation}

We can now compute the metric perturbations in the Poisson gauge 
using the transformation rules of ref. \cite{br96}. For the first 
order, we obtain
\begin{eqnarray}
\psi_P^{(1)}&=&\phi_P^{(1)}=\varphi,\nonumber\\
\chi_{P ij}^{(1)}&=&\chi_{ij}^{\top(1)}.
\end{eqnarray}
These equations show the well-known result for scalar perturbations 
in the longitudinal gauge and the gauge invariance for tensor 
modes at the linear level. For the second order, we obtain
\begin{eqnarray}
\psi_P^{(2)}&=&\eta^2\left(\frac{1}{6}\varphi^{,i}\varphi_{,i}
-\frac{10}{21}\Psi_0\right)+\frac{16}{3} \varphi^2+12  \Theta_0,
\nonumber\\
\phi_P^{(2)}&=&\eta^2\left(\frac{1}{6}\varphi^{,i}\varphi_{,i}
-\frac{10}{21}\Psi_0\right)+\frac{4}{3} \varphi^2-8  \Theta_0,
\nonumber\\
\nabla^2 z_P^{(2)i}&=&-\frac{8}{3}\eta\left(\varphi^{,i}
\nabla^2\varphi-\varphi^{,ij}\varphi_{,j}+2\Psi_0^{,i}\right),
\nonumber\\
\chi_{P ij}^{(2)}&=&\tilde{\pi}_{ij}.
\label{pg2}
\end{eqnarray}

Note that the resulting expressions for $\psi_P$ and $\phi_P$ can be 
recovered, except for the sub-leading time-independent terms, 
by taking the weak-field limit of Einstein's theory (e.g. ref. 
\cite{pjep93}) and then expanding in powers of the perturbation 
amplitude; this is basically the method employed in previous 
second order computations of the Rees-Sciama effect. 
Also interesting is the way in which the tensor modes appear 
in this gauge: the transformation from the synchronous to the Poisson 
gauge has in fact dropped the Newtonian and post-Newtonian 
contributions, whose physical interpretation in terms of 
gravitational waves is highly non-trivial (see the discussion 
in ref. \cite{mt96}); what remains is a wave-like piece plus a 
constant term which has no effects on $\delta T^{(2)}$. 

\section{Poisson gauge anisotropies}

Let us start by discussing the first order anisotropies that 
are described by eq. (\ref{dt1p}).
The first term $\psi^{(1)}_{\cal{E}}$ represents the contribution 
from the gravitational redshift of the photons due to the difference 
in gravitational potential between the emission and observation points.
$\psi^{(1)}_{\cal{O}}$ only contributes to the monopole and can
be neglected. The term $v^{(1)i}_{\cal{O}} e_i$ is the dipole due 
to the motion of the observer. The term $v^{(1)i}_{{\cal{E}}} e_i$
accounts for the Doppler effect due to the velocity of the photon-baryon 
fluid at recombination and contributes to the acoustic peaks.
The term $\tau$ describes the intrinsic anisotropies in the 
photon temperature and is highly model dependent. For example, for 
adiabatic perturbations, in which all the components (baryons, photons,
dark matter) have a constant number density ratio, the photon energy 
density, and thus the temperature, varies proportionally to the 
potential fluctuations (at scales larger than the Jeans length).
It can be seen  that in this case $\tau=\frac{1}{4}
\frac{\delta \rho_\gamma}{\rho_\gamma}|_{\cal{E}}=\frac{1}{3}
\frac{\delta \rho_T}{\rho_T}_{\cal{E}}=-\frac{2}{3}
\phi^{(1)}_{\cal{E}}$. It is the combination of this term 
and the first one that gives the standard result for adiabatic 
perturbations at large angular scales, $\delta T =\frac{1}{3} 
\varphi$. At small scales, $\tau$ gives the main contribution 
to the acoustic peaks. We have not intended here to include a 
computation of $\tau$ and $v^{(1)i}_{{\cal{E}}}$, that would involve 
solving the linearized transport equation for the photons, that is 
coupled to the fluid evolution equations for the cold dark matter 
component and the baryons, the Boltzmann equation for the neutrino 
distribution and the Einstein equations for the metric perturbations. 
This problem has been treated and solved numerically by several 
authors (see, e.g., \cite{pe70,bo84,vi84,bo87,ho89,su95,ma95}). 
We assume that $\tau$ and $v^{(1)i}_{{\cal{E}}}$ are known for a given 
model, and compute the additional anisotropy generated by the metric
perturbations along the photon path up to second order.

Finally, the contribution to $\delta T^{(1)}$ from the last 
term, $I_1(\lambda_{\cal{E}})$ (called integrated Sachs-Wolfe effect
and given by eq. (\ref{i1})), represents the additional gravitational 
redshift  due to the time variation of the metric during the photon 
travel.  As in the linear regime the scalar potentials $\phi$ and 
$\psi$ are constant in time for a flat matter dominated universe,
their contribution vanishes (they will 
however give a non-vanishing contribution in the non-linear regime). 
The contribution of tensor perturbations to the temperature 
anisotropies arises exclusively from $I_1(\lambda_{\cal{E}})$
 at the linear order.
In many inflationary models, in which besides the usual scalar 
perturbations also a background of gravitational waves is produced, 
their contribution to the CMB anisotropies can be comparable to that 
of scalar perturbations at large scales. The contribution from the 
observer's motion term is of order $10^{-3}$, while the remaining part 
contributes for an order $10^{-5}$. 

We can now discuss the second order anisotropies that are given
by eq. (\ref{dt2p}). The first term, given by
$\psi^{(2)}_{\cal{E}}$, represents the gravitational redshift 
of the photons due to the second order metric perturbations, and is 
much smaller than its first order equivalent.
Then, there are
several terms involving products of two of the terms contributing
to $\delta T^{(1)}$; these are all very small compared to 
$\delta T^{(1)}$ (at least three orders of magnitude smaller) and 
can safely  be neglected. Also the term 
$(\partial \tau/\partial d^i)d^{(1)i}$ is the product of two small 
quantities and can be neglected.

Then, there is the term $(x^{(1)j}_{\cal{E}}
+x^{(1)0}_{\cal{E}} e^j)\left(\psi^{(1)}_{,j}-v^{(1)}_{i,j} e^i+
\tau^{(1)}_{,j}\right)_{\cal{E}}$, that can be split into a piece 
proportional to
\begin{eqnarray}
x^{(1)j}_\bot (\lambda_{\cal{E}})
&\equiv& (\delta^{ij}-e^i e^j)x^{(1)}_i(\lambda_{\cal{E}})
\nonumber\\
&=&(\lambda_{\cal{E}}-\lambda_{\cal{O}}) 
\left(-\chi^{\top(1)jk}_{\cal{O}}e_k 
+\chi^{\top(1)ik}_{\cal{O}}e_k e_i e^j\right)
+\int_{\lambda_{\cal{O}}}^{\lambda_{\cal{E}}} d\lambda
\left(\chi^{\top(1)jk} e_k-\chi^{\top(1)ik} e_k e_i e^j\right)
\nonumber\\
&-&\int_{\lambda_{\cal{O}}}^{\lambda_{\cal{E}}} d\lambda
(\lambda_{\cal{E}}-\lambda)\left(2 \varphi^{,j}-2 \varphi_{,i}
e^i e^j-\frac{1}{2}\chi^{\top(1),j}_{kl}e^k e^l+
\frac{1}{2}\chi^{\top(1)}_{kl,i}e^k e^l e^i e^j\right),\label{glens}
\end{eqnarray}
and another piece proportional to
\begin{eqnarray}
(x^{(1)j}_\parallel+x^{(1)}_0 e^j)_{\cal{E}} 
&\equiv&e^j (x^{(1)}_i e^i+x^{(1)}_0)_{\cal{E}}
\nonumber\\
&=&-e^j\int_{\lambda_{\cal{O}}}^{\lambda_{\cal{E}}} d\lambda
\left(2 \varphi-\frac{1}{2}\chi^{\top(1)}_{kl}e^k e^l\right).
\label{tdel}
\end{eqnarray}
The first piece describes the effect of the gravitational lensing 
on the photons as they travel from the last scattering surface 
to the observer. The transverse displacement $x^{(1)j}_\bot$ includes 
the usual contribution from scalar perturbations ($\varphi$), 
that has been considered in some previous studies \cite{se96b,mu96} 
and has an observable effect on small angular scales, and a new 
contribution due to the gravitational wave background 
($\chi^{\top(1)}_{ij}$) acting as a source of lensing. The second 
piece is due to the time delay effect of the lenses that changes the 
spacelike distance to the intersection of the photon path with the
last scattering surface. The scalar part of this term is expected 
to be suppressed with respect to the gravitational lensing term
due to the spatial derivative of $\varphi$ that appears in eq.
(\ref{glens}). The gravitational waves part is probably of the same 
order of magnitude as its gravitational lensing counterpart.
The term $x^{(1)0}_{\cal{E}} A^{(1)'}_{\cal{E}}$ is similar in 
form to the gravitational lensing and time delay terms: it arises 
due to the difference in affine parameter along the perturbed and 
background geodesics. As it involves time derivatives of the metric 
perturbations, the contribution from scalar terms will be small, but 
the tensor contribution is expected to be larger.

The next term to consider couples the velocity of 
the photon-baryon fluid with the perturbation to the photon wave 
vector at emission and is given by $v^{(1)}_{{\cal{E}}i}
I_1^i(\lambda_{\cal{E}})$. 
Comparing it with the gravitational lensing contribution, 
we expect some reduction because the Doppler contribution to the 
first order anisotropies is smaller than the other first order terms
(although of the same order of magnitude) and  some enlargement because
of the factor $(\lambda_{\cal{E}}-\lambda)$ of difference between the   
$I_1^i(\lambda_{\cal{E}})$ and $x^{(1)j}_\bot$  expressions. 
A more careful quantitative estimate, which would require a choice of the 
particular structure formation model of interest, is beyond the aim of this 
paper. 

Finally, we have the term $I_2(\lambda_{\cal{E}})$, that is given by 
eq. (\ref{i2}) and is an integral of several terms. The first one 
accounts for the Rees-Sciama effect, given by
\begin{equation}
\delta T_{RS}=\frac{1}{2}\int_{\lambda_{\cal{O}}}^{\lambda_{\cal{E}}} 
d\lambda \left(\psi^{(2)'}+\phi^{(2)'}+z^{(2)'}_i e^i-
\frac{1}{2}\chi^{(2)'}_{ij}e^i e^j\right).\label{rs}
\end{equation}
We can use the second order perturbations of the metric obtained
in section III to compute it. The contribution from the scalar 
perturbations, $\psi^{(2)}$ and $\phi^{(2)}$, is given by
\begin{equation}
\delta T_{RS}=\int_{\eta_{\cal{O}}}^{\eta_{\cal{E}}} d\eta \eta
\left(\frac{1}{3}\varphi^{,i}\varphi_{,i}
-\frac{20}{21}\Psi_0\right),
\end{equation}
where the terms inside the brackets have to be evaluated along the 
background geodesic parametrized by $\lambda=\eta$.
This piece coincides with that considered in some previous studies 
of the Rees-Sciama effect \cite{ma90,ma92,ma94}. 
The resulting anisotropies 
turn out to be between one and two orders of magnitude smaller
that the first order ones. The contribution from the vector and 
tensor modes can be obtained by substituting $z^{(2)}_i$ and 
$\chi^{(2)}_{ij}$ from eq. (\ref{pg2}) into eq. (\ref{rs}). 
Let us estimate their magnitudes compared to that of the scalar piece. 
The integrand for the vector piece is $z^{(2)'}_i e^i \sim k \varphi^2$,
while the scalar one is $\psi^{(2)'} \sim \eta k^2 \varphi^2$.
Thus, $z^{(2)'}_i e^i \sim \psi^{(2)'}/(\eta k) \sim \psi^{(2)'}
(a H/k)$ and the vector contribution is suppressed with 
respect to the scalar one, as the wavelengths of interest are smaller 
than the Hubble radius. This estimate is similar to the one obtained
in ref. \cite{se96a}. 
In the same way, the integrand for the tensor piece is
$\chi^{(2)'}_{ij}e^i e^j \sim k \varphi^2/(k\eta)^2 \sim
\psi^{(2)'}/(k\eta)^3 \sim \psi^{(2)'}
(a H/k)^3$. Hence, also the tensor 
contribution is much suppressed with respect to the scalar one.

The integrand of the second term contributing to 
$I_2(\lambda_{\cal{E}})$ is $\chi^{\top(1)'}_{ij} e^j
(k^{(1)i}+e^i k^{(1)0})$, it represents a correction to the 
anisotropies generated by linear gravitational waves, due to the 
perturbation of the photon wave vector. The piece containing 
$k^{(1)0}$ is
expected to be smaller than the other one; the largest contribution 
can arise from the term
$$
-\int_{\lambda_{\cal{O}}}^{\lambda_{\cal{E}}} 
d\lambda \chi^{\top(1)'}_{ij} e^j \int_{\lambda_{\cal{O}}}^\lambda
d\lambda' A^{(1),i}.
$$

The last four terms in $I_2(\lambda_{\cal{E}})$ will have a 
small contribution coming from scalar perturbations as they involve 
time derivatives of the potentials $\psi$ and $\phi$ that are 
constant at linear order. The contribution coming from gravitational 
waves is expected to be larger, in particular the last two ones
$$
\int_{\lambda_{\cal{O}}}^{\lambda_{\cal{E}}} 
d\lambda \left(x^{(1)0} \chi^{\top(1)''}_{ij} e^i e^j+
x^{(1)k} \chi^{\top(1)'}_{ij,k} e^i e^j \right) . 
$$
These can be interpreted as the gravitational lensing and time delay 
effects acting on the anisotropies generated by the linear
gravitational wave background.

\section{Conclusions}

We have computed the anisotropies in the CMB radiation up to second 
order perturbations in the metric around a flat Robertson-Walker 
spacetime. This calculation generalizes the results of ref. 
\cite{py96} in that we have taken into account the velocity of the 
emitter and the observer, we have considered scalar, vector and tensor
perturbations and we have explicitly included the second order 
perturbations of the metric. We have obtained these second order 
metric perturbations for a universe filled with a collision-less 
fluid in the Poisson gauge, by performing a second order gauge 
transformation of the synchronous gauge solutions, that have already 
been studied in some detail in the literature. 

Using these results, we have discussed the relevance 
of the second order contributions to the anisotropies in the Poisson 
gauge. The most relevant expected contribution is due to the 
gravitational lensing of photons due to  density perturbations,
that has already been the subject of several studies. We have shown
that also a gravitational wave background acts as a source of lensing for the 
CMB photons. This effect is much smaller that the scalar one for 
a gravitational wave background with spectral index $n_T=0$ as
generated during an inflationary period. Other sources of 
gravitational waves with more power than the inflationary ones at 
small scales may give a larger contribution through this effect.
Other contributions
include the time delay effect of scalar and tensor lensing, a 
coupling of the velocity at emission with the perturbed photon wave
vector and a second order perturbation to the integrated Sachs-Wolfe 
piece. This term includes the well-known Rees-Sciama effect, that
has been widely studied for the time variation of the scalar
gravitational potential. Using the second order perturbed metric
in the Poisson gauge obtained in section III, we have shown that 
the additional contributions to the anisotropies arising from the 
vector and tensor modes induced by linear scalar perturbations
are expected to be suppressed with respect to the scalar one. We
have also pointed out the existence of two more terms that are 
corrections to the anisotropies generated by the linear gravitational 
wave background, due to the perturbation of the photon wave vector 
and to the lensing and time delay effects on gravitational wave 
anisotropies; these can give a relevant contribution to the 
integrated term. These contributions deserve a more detailed 
quantitative analysis. 

Although in the light of the present analysis 
we do not expect that the second order gravitational effects will 
give a major contribution to the anisotropies at any scale, it is 
interesting to know if they could be detected by the planned high 
accuracy satellite observations. The gravitational lensing
by scalar perturbations is known to give a few percent effect
in some structure formation models and thus will be relevant if the 
expected 1$\%$ sensitivity is achieved.
The amplitude of the second order 
terms is also important because they contribute to the theoretical 
error of the anisotropy computations that will be used to determine 
the cosmological parameters from the measured multipoles.

\acknowledgements
S. Mollerach acknowledges the Vicerrectorado de investigaci\'on
de la Universidad de Valencia for financial support, 
S. Matarrese acknowledges the Italian MURST for
partial financial support.

\end{document}